# Phase Stability of Multiferroic GaFeO$_3$ up to 1368 K from In situ Neutron Diffraction

S. K. Mishra, R. Mittal and R. Singh

Solid State Physics Division, Bhabha Atomic Research Centre, Trombay, Mumbai-400085, India

M. Zbiri, T. Hansen and H. Schober

Institut Laue-Langevin, BP 156, 38042 Grenoble Cedex 9, France

We report a detailed high-temperature powder neutron diffraction investigation of the structural behavior of the multiferroic GaFeO$_3$ between 296 and 1368 K. Temperature dependent neutron diffraction patterns do not show any appreciable change either in intensity or appearance/disappearance of the observed peaks upto 1368 K, ruling out any structural transition in the entire temperature range. The lattice parameters and volume exhibit normal thermal expansion behaviour, indicating the absence of any structural changes with increasing temperature. The origin of the magnetoelectric couplings and multiferroicity in GaFeO$_3$ is known to be influenced by the site disorder from Ga/Fe atoms. Our analysis shows that this disorder remains nearly the same upon increase of temperature from 296 to 1368 K. The structural parameters as obtained from Rietveld refinement of neutron diffraction data are used to calculate the interatomic distances and distortions of the oxygen polyhedra around the Ga$_1$, Ga$_2$, Fe$_1$ and Fe$_2$ cations. Evolution of the distortion of the oxygen polyhedra around these sites suggests that the Ga$_1$-O tetrahedron is least distorted and Fe$_1$-O is most distorted. Structural features regarding the distortion of polyhedral units would be crucial to understand the temperature dependence of the microscopic origin of polarizations. The electric polarization has been estimated using a simple ionic model and its value is found to decrease with increasing temperature.

# Introduction

Functional oxides represent a broad class of materials that exhibit a wide range of crystal structures and functionalities [1-4]. Among these, detailed study of properties associated with ferroic interactions like antiferro/ferro electric, antiferro/ferro magnetic and more recently the combination of these two known as multiferroic has stimulated considerable attention of the scientific community [1-4]. The studies of these properties have experienced a dramatic increase in research efforts over the past decade. Based on these efforts, multiferroics emerge for real life applications in microelectronics, spintronics, data storage, and computing hardware industry [1,4]. At present, there are a few materials, which possess multiferroicity due to exclusion of electrical and magnetic spontaneous polarization mutually. $BiFeO_3$ or $LuFe_2O_4$ and $RMnO_3$ (R = rare earth) show multiferroicity due to covalent bonding, electronic and geometric ferroelectricities [5-8]. $GaFeO_3$ is also a prominent candidate of mutiferroic materials due to its large magnetoelectric effect, magneto-optic, and piezoelectric properties [9-28].

$GaFeO_3$ has been investigated by using various techniques, which show interesting properties of this material such as magnetization-induced second harmonic generation [13], optical and dc magnetoelectric effect [12,14], as well as ultrafast electric and magnetic response induced by irradiation of a femtosecond laser pulse [15], Faraday rotation [16]. The compound was also found to exhibit an unusually large orbital magnetic moment [17]. These properties make this compound very attractive for potential applications.

At ambient condition, $GaFeO_3$ possesses a noncentrosymmetric orthorhombic structure with space group $Pc2_1n$, which indicates that the spontaneous polarization is along the *b* axis in the paramagnetic phase. Upon cooling, it undergoes a magnetic transition from a paramagnetic to a ferri-magnetic phase at around 225 K. The ferri-magnetism is due to the unequal distribution of Fe spins of nearly equal magnitude on the sub-lattices with a magnetic moment of the spin along the c-axis [18]. The origin of the unequal distribution of Fe spins in $GaFeO_3$ is due to the Ga-Fe disorder which makes this compound interesting. The relationship between the magnetic moment and Ga-Fe disorder has been studied theoretically [27]. Conjunction of magnetic susceptibility, magnetization and ESR experiments confirmed that $GaFeO_3$ is not a classical ferrimagnet (the $Fe^{3+}$ spin system can-not be divided into two sublattices) [28]. It can be qualitatively explained by considering three sublattices and which are related to three nonequivalent cationic octahedral sites $Fe_1$, $Fe_2$, and $Ga_2$ respectively. The origin of magnetoelectric couplings and multiferroicity appears to be influenced by the disorder nature of Ga/Fe.

Detailed structural studies are essential for understanding the mechanism of the magnetoelectric coupling which would lead to deeper insights into structure property correlations.

A number of structural studies on GaFeO$_3$ have been reported in the literature using X-ray, synchrotron and neutron diffraction techniques below room temperature and reveal no structural phase transition. In conjunction with high temperature powder x-ray diffraction and Raman scattering study, Mukherjee *et al.* [24] showed the absence of structural phase transitions in GaFeO$_3$ up to 700 K. The first estimation of the electric polarization using neutron diffraction data was done by Arima et al [12], which gives a value of $P_S \approx 2.5$ µC/cm$^2$. The computed spontaneous polarization of GaFeO$_3$ using first principle calculation [25] has been found to be 59 µC/cm$^2$ from the structure obtained from the minimization of the free energy. Recently, D. Stoeffer [26] has predicted the polarization in the multiferroic GaFeO$_3$ to be $P_S \approx 25$ µC/cm$^2$ from the calculated electronic structure using first principles methods. The emergence of ferroelectricity is due to tilting and distortion of FeO$_6$ octahedra and the off-centering shift of the Fe ions. Interestingly, it is more appropriate to consider it as ferrielectricity since it is due to opposite dipoles induced by opposite but unequal displacements and disorder of the Ga/Fe ions. In general, ferroelectric materials undergo a structural phase transition that is paralectric in nature at the highest temperatures and that is accompanied with anomalies in physical properties such as the dielectric constant. There are experimental difficulties in measuring dielectric constants at high temperatures due to a large contribution of conductivity. Therefore, a systematic temperature dependent study with careful examination of the distortion of the structure is required.

In the present study, we have carried out systematic temperature-dependent neutron diffraction measurements from 296 to 1368 K. Structure evolution, polyhedral distortion and structural parameters were investigated as a function of temperature. Neutron diffraction offers certain advantages over X-rays especially in the accurate determination of the oxygen positions that are crucial for computing polyhedral distortion and that are also important for ferroelectricity. The interpretation of the magnetic structure and other properties requires prior determination of the occupancies of the cationic sites. Neutron diffraction also provide accurate information about the distribution of these sites thanks to the good neutron scattering contrast between Ga$^{3+}$ and Fe$^{3+}$ compared to x-rays, which offer a small scattering contrast. Therefore, this advantage has been used to elucidate temperature-dependent structural changes, and site specific information. Presently no signature of a structural phase transition up to 1368 K was found. The polarization has been estimated using a simple ionic model, which is found to be in fair agreement with the value reported using first principles [26], and decreases with increasing temperature.

**Experimental**

GaFeO$_3$ polycrystalline samples were prepared by the solid-state reaction method. The powder x-ray diffraction measurement at ambient conditions confirmed the single-phase nature of the samples. The temperature dependent neutron powder diffraction experiments were performed on the high-flux D20 diffractometer at the Institut Laue- Langevin, (Grenoble, France) [29]. The "high resolution" mode (take-off angle 118º) was selected with a wavelength of 1.3594 Å (vertically focusing Germanium monochromator in reflection, (117) reflection). The sample was heated up to 1368 K in a quartz tube. The other end of the quartz tube was open to air to ensure keeping the initial oxygen stoichiometry. The structural refinements were performed using the Rietveld method within the program FULLPROF [30]. In all the refinements, the background was defined by a point to point in 2θ. A Thompson-Cox-Hastings pseudo-Voigt with axial divergence asymmetry function was chosen to define the profile shape for the neutron diffraction peaks. All other parameters i.e., scale factor, zero correction, background and half-width parameters along with mixing parameters, lattice parameters, positional coordinates, and thermal parameters were refined. In all the refinements the data over the full angular range of 10≤2θ≤ 130 degree was used; although in various figures only a limited range is shown, for the sake of clarity.

**Results and Discussion**

Fig. 1 depicts a portion of the powder neutron diffraction patterns of GaFeO$_3$ as a function of temperature in the range 296 – 1368 K, during the heating cycle. It is evident that the temperature dependent neutron diffraction patterns do not show any appreciable change neither in the intensity nor the appearance or disappearance of any reflections up to 1368 K. This clearly suggests absence of any structural change in the entire temperature range. At ambient conditions, the system crystallizes in a paramagnetic phase with a non-centrosymmetric orthorhombic structure having the space group $Pc2_1n$ [9-12]. This structural phase has eight formula units per unit-cell with four inequivalent cationic sites: Ga$_1$ ions are tetrahedrally coordinated while Ga$_2$, Fe$_1$ and Fe$_2$ ions occupying all the octahedral sites as shown in Fig. 2.

Due to the isovalency and the close ionic radii of Ga and Fe (0.62 and 0.65 Å, respectively) site disordering in GaFeO$_3$ is expected. In order to determine the cationic site occupancies (disorder), Ga and Fe cations were introduced at Ga$_1$, Ga$_2$ and Fe$_1$ and Fe$_2$ positions respectively, and the occupancy factors were refined, constrained to full occupancy, notably improving the quality of the fit. Then, the possibility of partial occupancy of Ga$_1$ and Ga$_2$ positions by some Fe cations and *vice versa* was also checked and this also led to a drop of the discrepancy factor ($\chi^2$). Finally, we refined the cation occupancies along with lattice parameters, positional coordinates and thermal parameters. However, anionic sites were kept fully occupied. The obtained results indicate that the Ga$_2$ site (0.36) is occupied predominantly by Fe and as compared to Ga$_1$ site (0.02). Band structure calculations [27] also indicate that the energy resulting from the Fe interchange with the Ga$_2$ site can be as small as 1 meV per formula unit while the interchange with a Ga$_1$ site requires an energy almost two orders of magnitude larger. This explains why Ga$_2$-Fe disorder is high. The occupancies of iron (Fe) at Ga$_1$, Ga$_2$, Fe$_1$ and Fe$_2$ sites are found to be 0.02, 0.36, 0.88, 0.63 at 296 K and 0.07, 0.37, 0.87 and 0.62 at 1368 K, respectively. This suggests that there is no significant change in disorder as a function of temperature. Thus site occupancies for cations were kept constant at the values corresponding to 296 K in the subsequent refinements for higher temperatures.

Detailed Rietveld refinement of the powder diffraction data shows that the temperature dependent neutron diffraction patterns could be indexed using the same orthorhombic structure up to 1368 K. The Rietveld refinements were accomplished smoothly, revealing a monotonic increase in the lattice constant and cell volume with increasing temperature. The fit between the observed and calculated profiles is satisfactory and some of them are shown in (Fig. 3 (a-d)). The smooth variation of the lattice parameters, thermal displacement factors with temperature in the entire high-temperature regime confirms that there is no high-temperature phase transition. The detailed structural parameters and goodness of fit at selected temperatures, as obtained from neutron diffraction data, are given in Table I.

Evolution of the lattice parameters and the unit cell volume as obtained from the refinement of the neutron diffraction data are shown in Fig. 4. All lattice parameters and volume exhibit regular thermal expansion behaviour, indicating no abnormal structural changes with increasing temperature. From the linear fitting of lattice parameters, the coefficient of thermal expansion ($\alpha$/K) of GaFeO$_3$ along <100>, <010> and <001> (i.e the principle crystallographic axes) were found to have values of $9.56 \times 10^{-6}$, $9.40 \times 10^{-6}$ and $9.30 \times 10^{-6}$, respectively. The small difference among these three values suggests that GaFeO$_3$ is a thermally isotropic material. The coefficient of volume thermal expansion is found to be $28.7 \times 10^{-6}$ K$^{-1}$. This is in good agreement with the relation $\gamma = 3\alpha$, satisfied by isotropic materials.

The variation of the isotropic thermal parameters for all the atoms in the asymmetric unit cell is shown in Figure 5. It is evident that the thermal parameters of all the atoms increase with increasing temperature up to 1050 K followed by an apparent anomalous increase for O$_3$ and decrease for O$_4$ and O$_5$. The B$_{iso}$ for O$_3$ atoms shows a large enhancement from 1.7 Å$^2$ (at 1072 K) to 3.45Å$^2$ (at 1368 K). This might suggest appearance of a soft mode. However, it should be noted that on fixing the thermal parameter of O$_3$ around B$_{iso}$= 2.45, the refinement worsens only slightly ($\chi^2$ increases from 4.19 to 4.36) and this does not significantly change other thermal parameters and the position parameters. Large increase in thermal parameters of all the oxygens was also observed in YMnO$_3$ at ~ 920 K prior to phase transition at 1273 K [31]. Structural parameters are used to estimate the variation of characteristic mean interatomic distance (mean bond length) with temperature (Fig. 6). The variation of bond lengths suggests that the octahedral coordination for both Fe$_1$ and Fe$_2$ sites have highly distorted Fe–O bonds. However the tetrahedral coordination around the Ga$_1$ site is quite regular. The coordination polyhedron around Ga$_2$ is significantly smaller in comparison with those around Fe$_1$ and Fe$_2$ sites. All bond lengths are found to show a monotonic increase with temperature within the experimental error. The Ga$_1$-O bond lengths show weak anomalous behavior around 900 K and Ga$_2$-O$_1$ bond lengths show regressive trends above 1100 K. It should be noted that we have not found any contraction of the Fe-O bond-lengths within the entire explored temperature range, contrary to what was observed by S. Mukherjee et al [24] using x-ray diffraction. This discrepancy may be due to the fact that neutron diffraction has certain advantages over X-ray, especially in the accurate determination of oxygen positions.

The macroscopic properties of the sample linked to a short range ordering of the cations which cause specific local distortions of the oxygen polyhedra around the cations. Thus, we have estimated the distortion of the polyhedra as a function of temperature by using structural parameters obtained from Rietveld refinements of neutron diffraction data using Fullprof (Bond Str sub-programme) software [30]. Fig. 7 shows the evolution of the distortions of the oxygen polyhedra around the Ga$_1$, Ga$_2$, Fe$_1$ and Fe$_2$ cations. It comes out that the Ga$_1$-O tetrahedron is the least distorted and the Fe$_1$-O is the most distorted. This suggests that the Ga$_1$-O tetrahedron is naturally regular. Upon increasing temperature to 1368 K, the distortion increases for the Ga$_1$-O tetrahedron and the Fe$_2$-O octahedron, but decreases for the Ga$_2$-O octahedron up to 1100 K. Above, it becomes nearly temperature independent. On the other hand the distortion for the Fe$_1$-O octahedron is found to increase in the entire temperature range. Such structural features regarding the distortion of GaFeO$_3$ would be crucial to understand the temperature dependence of the microscopic origin of polarizations in the sample.

The estimation of spontaneous polarization ($P_S$) in GaFeO$_3$ was computed by various authors using different approaches. Based on the off-center displacement of Fe ions in FeO$_6$ octahedra, Arima *et al* [12] predicted a $P_S \approx 2.5\ \mu C/cm^2$. But, such a point charge calculation does not provide a correct estimate of $Ps$, since various other contributions from the structural features such as the Ga$_1$–O tetrahedra and Ga$_2$–O octahedra, were neglected. Recently, D. Stoeffer [26] predicted the value of the polarization of GaFeO$_3$ to be $P_S \approx 25\ \mu C/cm^2$, which is ten times larger than the previous calculation by Arima *et al* [12]. This value was based on estimating the electronic structure using first principles methods and considering the modern theory of polarization. A Roy *et al.* [25] also calculated the $Ps$ of GaFeO$_3$ in its ground state using both nominal ionic charges and calculated Born effective charges. The nominal ionic charges calculation yielded $Ps$ of $\sim 30.53\ \mu C/cm^2$, almost half the value obtained using the Born effective charges ($Ps$ of $\sim 59\ \mu C/cm^2$). The authors concluded that the spontaneous polarization in GaFeO$_3$ is primarily contributed by the asymmetrically located Ga$_1$, Fe$_1$, O$_1$, O$_2$ and O$_6$ ions. However, at elevated temperatures, site disordering between Fe$_1$ and Ga$_1$ sites substantially lower the spontaneous polarization.

In the present study, the polarization has been estimated from the refined structures using a simple ionic model $P = c_i Q_i e m_i / V$, where $c_i$ is the displacement of the site from the centrosymmetric position in Å, $Q_i$ is the ionic charge, $e$ is the electron charge, the site multiplicity is denoted by $m_i$ and the unit cell volume is denoted by $V$. The results of this estimate are shown in Fig. 8. We have also estimated the $Ps\ value$, using the contributions from all the structural sources such as Ga$_1$–O tetrahedra, Ga$_2$–O, Fe$_1$–O and Fe$_2$–O octahedra. The value of the spontaneous polarization is found to decrease with increasing temperature, from ~15.3 $\mu C/cm^2$ (at 296 K) to 14.0 $\mu C/cm^2$ (at 1368 K) along the direction <010>. This could be easily explained from a crystallographic point of view. GaFeO$_3$ has orthorhombic structure with the *Pc2$_1$n* space group. The first and third operations (*c* and *n* respectively) on the atom positions do not put any constraint on the corresponding displacement and in turn the polarization vector remains unrestricted. But, the 2$_1$ symmetry operator (screw rotated by 180° about the [010]-axis), changes the atomic positions from (*x, y, z*) to (−*x, y,*−*z*). As a result of this, the crystal polarization changes from (*Px , Py , Pz*) to (−*Px , Py ,*−*Pz*) leading to a zero crystal polarization along the *a*- and *c*-axes and non-zero along the *b*-axis. The estimated polarization value is found to be in a fairly good agreement with the one reported using first principle calculations [26].

We have computed the charge density using the structural parameters obtained from the Rietveld refinement of neutron diffraction data and the VESTA software [32]. This is shown in Figure 9. The charge density can shed light on bonding in $GaFeO_3$, especially the partial covalency character of the cation–anion bonds, which can be further correlated with the functional properties of $GaFeO_3$. Our findings suggest that although most of the charges are symmetrically distributed along the radius of the atomic circles, indicating the largely ionic nature of bonding, a small amount of covalency is shown by minor asymmetry of the charges around the oxygen anions connected to the $Fe_1$, $Fe_2$, $Ga_1$ and $Ga_2$ ions.

## Conclusions

In conclusion, we have carried out a systematic neutron diffraction measurements of the structural parameters of the multiferroic material $GaFeO_3$, in the temperature range 296 - 1368 K. The results do not show any appreciable change in diffraction patterns as a function of temperature which rules out any structural phase transition in the entire explored temperature domain. We found that the Ga/Fe site disorder remains nearly un-change upon increase of temperature from 296 to1368 K. This might be crucial for the understanding of the origin of magnetoelectric coupling and multiferroicity in $GaFeO_3$. All the lattice parameters and the volume exhibit regular thermal expansion behaviour, and confirm the absence of any structural changes with increasing temperature. Structural parameters are used to calculate bond lengths and the distortions of the oxygen polyhedra around the $Ga_1$, $Ga_2$, $Fe_1$ and $Fe_2$ cations. The polarization has been estimated using a simple ionic model and found to be in good agreement with the value reported using first principle [26]. We have found spontaneous polarization whose value decreases with increasing temperature from ~15.3 $\mu C/cm^2$ (at 296 K) to 14.0 $\mu C/cm^2$ (at 1368 K).

1. M. E. Lines and A. M. Glass, *Principles and Application of Ferroelectrics and Related Materials* (Clarendon, Oxford, 1977); L. G. Tejuca and J. L. G. Fierro, *Properties and Applications of Perovskite-Type Oxides* (Dekker, New York, 1993).
2. M. Bibes and A. Barthelemy, Nature Mater. **7,** 425 (2008); T. Kimura, Y. Sekio, H. Nakamura, T. Siegrist and A. P. Ramirez, Nature Mater. **7,** 291 (2008); M. Mostovoy, Nature Mater. **7,** 269 (2008); Y. H. Chu, L W Martin, M B Holcomb, M Gajek, S-J Han, Q He, N balke, C h Yang, D Lee, W Hu, Q Zhan, Pei Ling Yang, A F Rodriguez, A Scholl, S X Wang and R Ramesh, Nature Mater. **7,** 478 (2008).
3. R. Ramesh and N. A. Spaldin, Nature mater. **6,** 21 (2007), S. W. Cheong and M. Mostovory, Nature Mater. **6,** 13 (2007); W. Eerenstein, N. D. Mathur, and J.F. Scott, Nature (*London*) **442**, 759 (2006).
4. E. Bousquet, M. Dawber, N. Stucki, C. Lichtensteiger, P. Hermet, S. Gariglio, J. M. Triscone, and P. Ghosez, Nature (London) **452**, 723 (2008); T. Choi, Y. Horibe, H. T. Yi, Y. J. Choi, WeidaWu, and S.-W. Cheong, Nature Mater. **9**, 253 (2010).
5. T. Kimura, S. Kawamoto, I. Yamada, M. Azuma, M. Takano, and Y. Tokura, Phys. Rev. B **67**, 180401(R) (2003).
6. Naoshi Ikeda, Hiroyuki Ohsumi, Kenji Ohwada, Kenji Ishii, Toshiya Inami, Kazuhisa Kakurai,YouichiMurakami,KenjiYoshii, Shigeo Mori, Yoichi Horibe, and Hijiri Kito, Nature (London) **436**, 1136 (2005); F. M. Vitucci, A. Nucara, D. Nicoletti, Y. Sun, C. H. Li, J. C. Soret, U. Schade, and P. Calvani, Phys. Rev. B **81**, 195121 (2010).
7. T. Katsufuji, S. Mori, M. Masaki, Y. Moritomo, N. Yamamoto, and H. Takagi, Phys. Rev. B **64**, 104419 (2001).
8. M. Zbiri, H. Schober, N. Choudhury, R. Mittal, S. Chaplot, S. Patwe, S. Achary, and A. Tyagi, Appl. Phys. Lett. **100**, 142901 (2012).
9. J. P. Remeika, J. Appl. Phys. Suppl. **31**, 263S (1960).
10. S. C. Abrahams and J. M. Reddy, Phys. Rev. Lett. **13**, 688 (1964);
11. George T. Rado, Phys. Rev. Lett. **13**, 335 (1964).
12. T. Arima, D. Higashiyama, Y. Kaneko, J. P. He, T. Goto, S. Miyasaka, T. Kimura, K. Oikawa, T. Kamiyama, R. Kumai and Y. Tokura, Phys. Rev. B **70**, 064426 (2004).
13. Y. Ogawa, Y. Kaneko, J. P. He, X. Z. Yu, T. Arima, and Y. Tokura, Phys. Rev. Lett. **92**, 047401 (2004); A. M. Kalashnikova, R. V. Pisarev, L. N. Bezmaternykh, V. L. Temerov, A. Kirilyuk, and Th. Rasing, JETP Lett. **81**, 452 (2005); Jun-ichi Igarashi and Tatsuya Nagao, Phys. Rev B **82**, 024424 (2010).
14. J H Jung, M Matubara, T Arima, J P He, Y Kaneko and Y Tokura Phys. Rev. Lett. **93**, 037403 (2004); M. Kubota, T. Arima, Y. Kaneko, J. P. He, X. Z. Yu, and Y. Tokura, Phys. Rev. Lett. **92**, 137401 (2004); N. Kida, Y. Kaneko, J. P. He, M. Matsubara, H. Sato, T. Arima, H. Akoh and Y. Tokura, Phys. Rev. Lett. **96**, 167202 (2006).

15. M. Matsubara, Y. Kaneko, Jin-Ping He, H. Okamoto, and Y. Tokura, Phys. Rev B **79**, 140411 (R) (2009).
16. A. M. Kalashnikova, R. V. Pisarev, L. N. Bezmaternykh, V. L. Temerov, A. Kirilyuk and Th. Rasing, JETP Lett. **81**, 452 (2005).
17. J-Y Kim, T Y Koo and J-H Park Phys. Rev. Lett. **96**, 047205 (2006).
18. R. B. Frankel, N. A. Blum, S. Foner, A. J. Freeman and M. Schieber, Phys. Rev. Lett. **15**, 958 (1965).
19. Yu. F. Popov, A. M. Kadomtseva, G. P. Vorob'ev, V. A. Timofeeva, D. M. Ustinin, A. K. Zvezdin, and M. M. Tegeranchi, Zh. Eksp. Teor. Fiz. **114**, 263 (1998) [JETP **87**, 146 (1998)].
20. Stephen W. Lovesey, Kevin S. Knight, and Ewald Balcar, J. Phys.: Condens. Matter **19**, 376205 (2007).
21. V. B. Naik and R. Mahendiran, J. Appl. Phys. **106**, 123910 (2009).
22. U. Staub, Y. Bodenthin, C. Piamonteze, S. P. Collins, S. Koohpayeh, D. Fort, and S. W. Lovesey, Phys. Rev. B **82**, 104411 (2010).
23. A. Shireen, R. Saha, P. Mandal, A. Sundaresan, and C. N. R. Rao, J. Mater. Chem. **21**, 57 (2011).
24. S. Mukherjee, A. Garg and R. Gupta, J. Phys.: Condens. Matter **23**, 445403 (2011).
25. A. Roy, S Mukherjee, R Gupta, S. Auluck, R. Prasad and A. Garg, J. Phys.: Condens. Matter **23**, 325902 (2011).
26. D Stoeffler, J. Phys.: Condens. Matter **24**, 185502 (2012).
27. M. J. Han, T. Ozaki, and J. Yu, Phys. Rev B. **75**, 060404 (2007).
28. G. Gruener, F. Vitucci, P. Calvani and J. C. Soret, Phys. Rev B. **84**, 224427 (2011).
29. T. Hansen, P.F. Henry, H.E. Fischer, J. Torregrossa, P. Convert, Meas Sci Technol **19**, 034001 (2008).
30. J. Rodriguez-Carvajal Physica **B 192**, 55 (1993).
31. A. S. Gibbs, K. S. Knight and P. Lightfoot, Phys. Rev. B **83**, 094111 (2011).
32. K. Momma and F. Izumi, J. Appl. Crystallogr. **41**, 653 (2008).

**Table I:** Structural parameters of GaFeO$_3$ obtained by Rietveld refinement of neutron diffraction data collected at 677 K and 1368 K, using *Pc2$_1$n* space group. The occupancies of iron at Ga$_1$, Ga$_2$, Fe$_1$ and Fe$_2$ sites are 0.02, 0.36, 0.88, 0.63 at 677 K, and 0.07, 0.37, 0.87 and 0.62 at 1368 K, respectively.

| | Temperature = 677 K | | | | | Temperature = 1368 K | | | |
|---|---|---|---|---|---|---|---|---|---|
| **Atoms Positional Coordinates and thermal parameter** | | | | | **Atoms Positional Coordinates and thermal parameter** | | | | |
| | **X** | **Y** | **Z** | **B(Å)$^2$** | | **X** | **Y** | **Z** | **B(Å)$^2$** |
| **Ga$_1$** | 0.1536(9) | 0.0000 | 0.1755(2) | 0.89(8) | **Ga$_1$** | 0.1518(5) | 0.0000 | 0.1749(5) | 1.59(5) |
| **Ga$_2$** | 0.1560(7) | 0.3076(7) | 0.8121(8) | 0.79(6) | **Ga$_2$** | 0.1587(6) | 0.3061(5) | 0.8134(8) | 1.71(3) |
| **Fe$_1$** | 0.1515(6) | 0.5849(6) | 0.1857(9) | 0.86(5) | **Fe$_1$** | 0.1496(6) | 0.5850(6) | 0.1865(9) | 1.91(3) |
| **Fe$_2$** | 0.0346(5) | 0.7958(8) | 0.6772(6) | 0.92(5) | **Fe$_2$** | 0.0347(5) | 0.7942(8) | 0.6774(7) | 1.82(3) |
| **O1** | 0.3246(6) | 0.4236(9) | 0.9803(5) | 1.18(8) | **O1** | 0.3241(6) | 0.4195(9) | 0.9830(8) | 2.35(7) |
| **O2** | 0.4900(1) | 0.4354(6) | 0.5157(8) | 0.65(6) | **O2** | 0.4912(3) | 0.4369(6) | 0.5154(7) | 1.34(6) |
| **O3** | 0.9962(2) | 0.2024(4) | 0.6541(5) | 0.90(9) | **O3** | 0.9897(5) | 0.2035(6) | 0.6578(6) | 3.45(8) |
| **O4** | 0.1574(5) | 0.1957(6) | 0.1637(7) | 1.01(8) | **O4** | 0.1641(5) | 0.1939(7) | 0.1621(7) | 1.60(7) |
| **O5** | 0.1698(6) | 0.6699(4) | 0.8482(5) | 0.95(8) | **O5** | 0.1741(4) | 0.6695(6) | 0.8503(6) | 1.16(6) |
| **O6** | 0.1691(5) | 0.9344(6) | 0.5176(7) | 1.03(7) | **O6** | 0.1670(6) | 0.9334(5) | 0.5213(7) | 2.39(5) |
| Lattice Parameters (Å) | | | | | Lattice Parameters (Å) | | | | |
| A= 8.7681 (1) (Å); B= 9.4223 (1) (Å) | | | | | A= 8.8298(1) (Å); B= 9.4865(2) (Å) | | | | |
| C= 5.0999 (1) (Å); Volume = 421.33(4) (Å)$^3$ | | | | | C= 5.1333(1) (Å); Volume = 429.986(6) (Å)$^3$ | | | | |
| R$_p$=7.32; R$_{wp}$=8.28; R$_{exp}$=2.96; $\chi^2$= 7.85 | | | | | R$_p$=9.21; R$_{wp}$=9.36; R$_{exp}$=4.57; $\chi^2$= 4.19 | | | | |

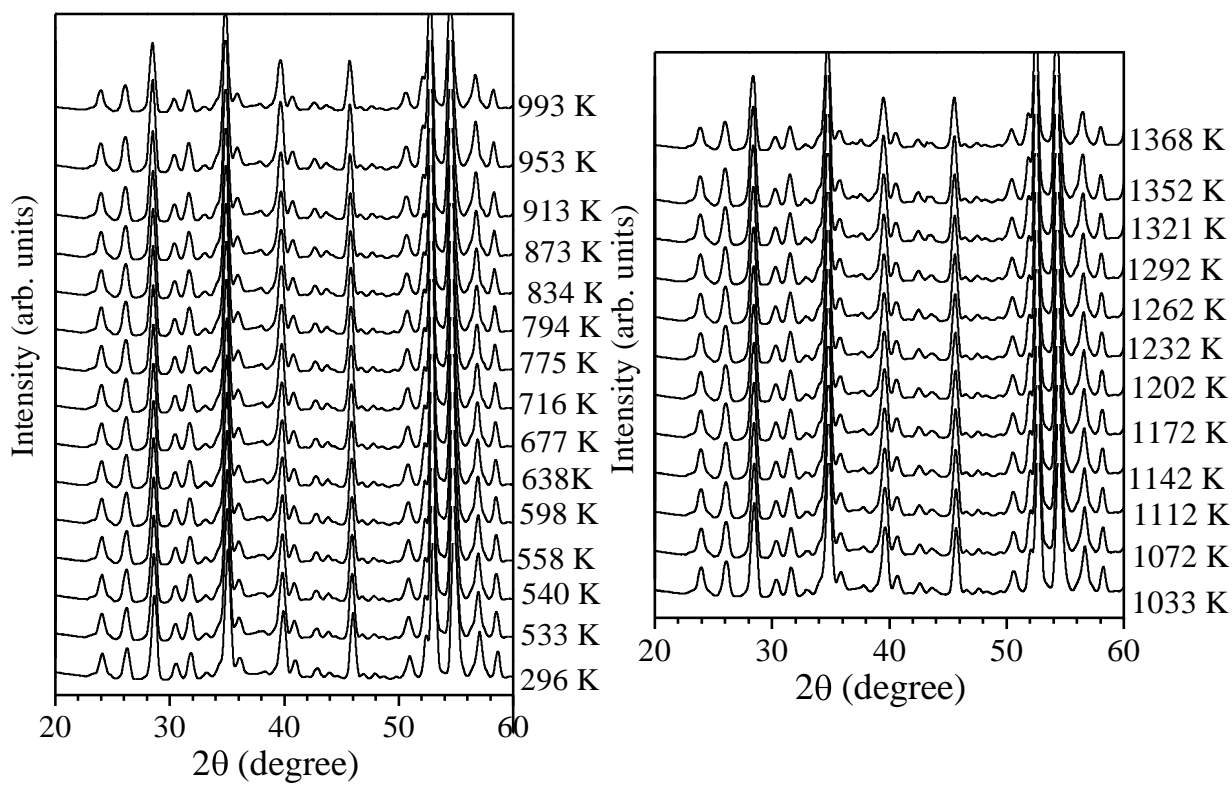

**Fig. 1** Evolution of the neutron diffraction patterns for GaFeO$_3$ as a function of temperature.

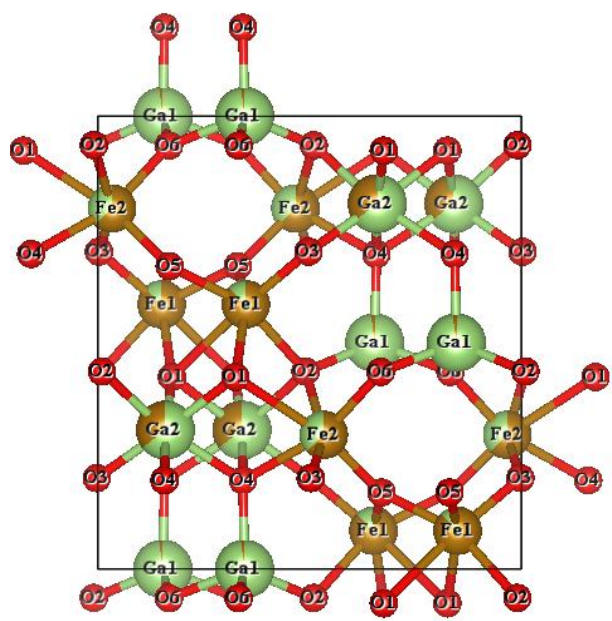

**Fig. 2** Crystal structure of GaFeO$_3$ at room temperature.

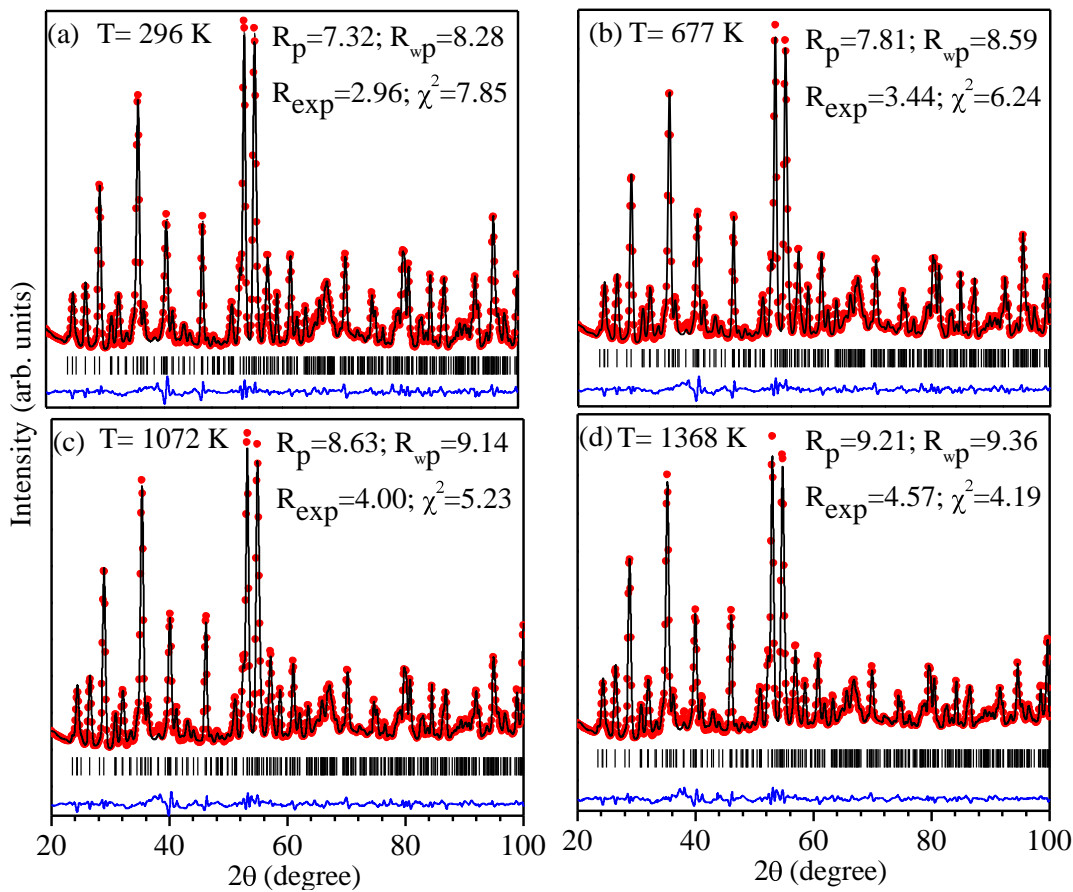

**Fig. 3** (color online) Observed (circle), calculated (continuous line), and difference (bottom line) profiles obtained after the Rietveld refinement of GaFeO$_3$ using the orthorhombic space group $Pc2_1n$ at temperatures: (a) 296 K, (b) 677 K, (c) 1072 K and (d) 1368 K, respectively.

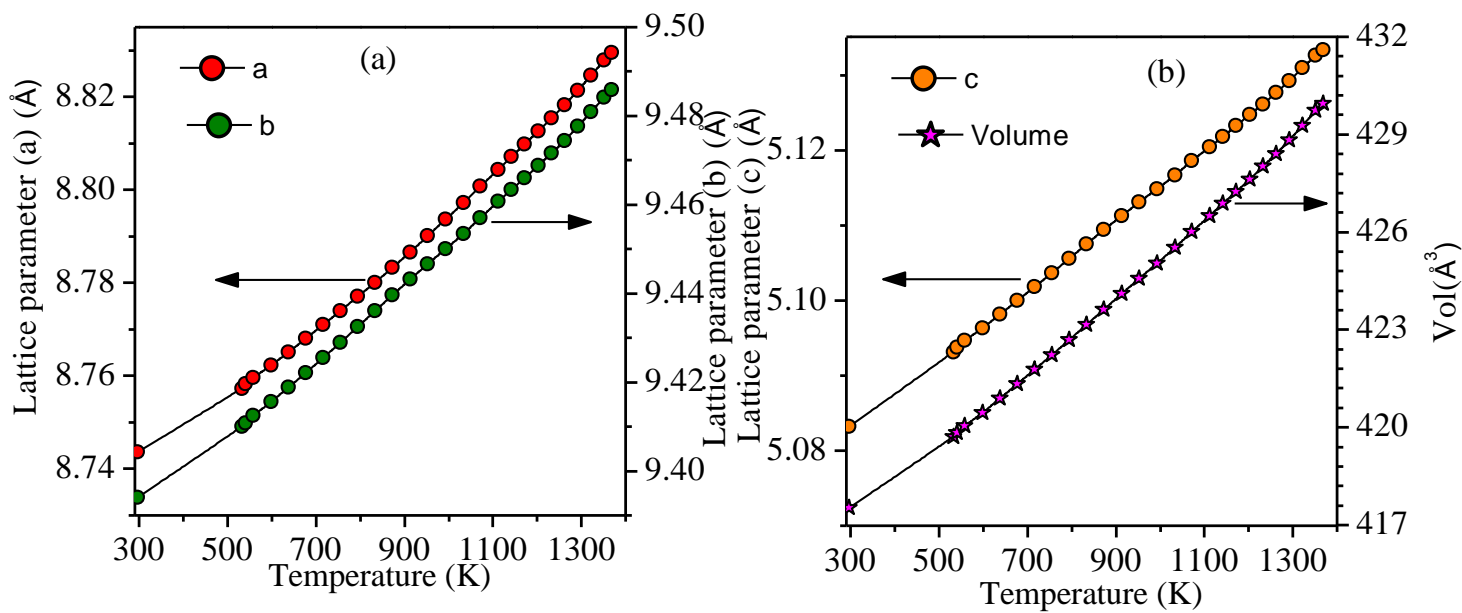

**Fig. 4** (color online) The variation of the lattice parameters and volume as a function of temperature obtained after Rietveld refinement for GaFeO$_3$.

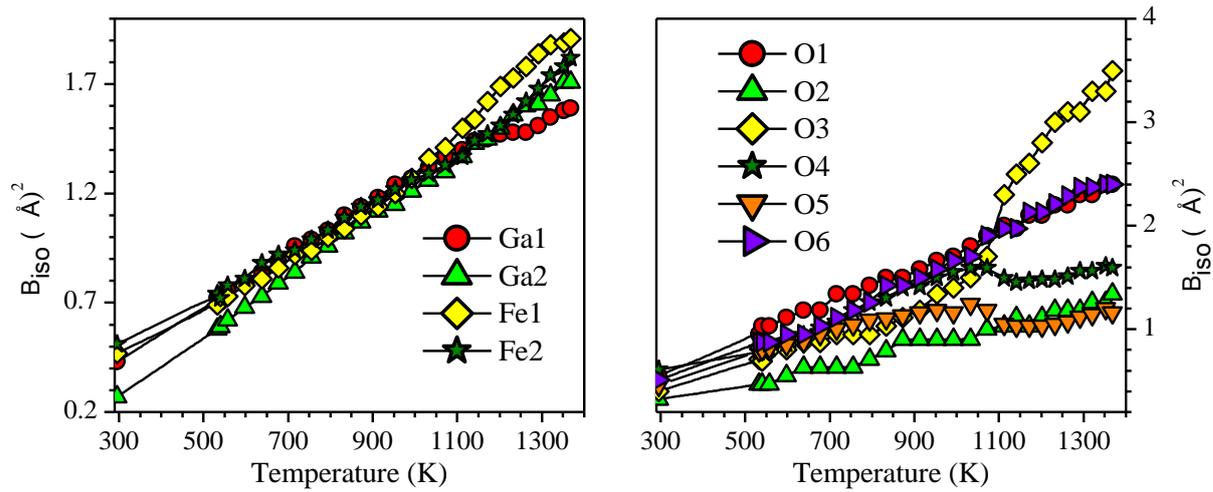

**Fig. 5** (color online) Evolution of the isotropic thermal parameters as a function of temperature obtained after Rietveld refinement for GaFeO$_3$.

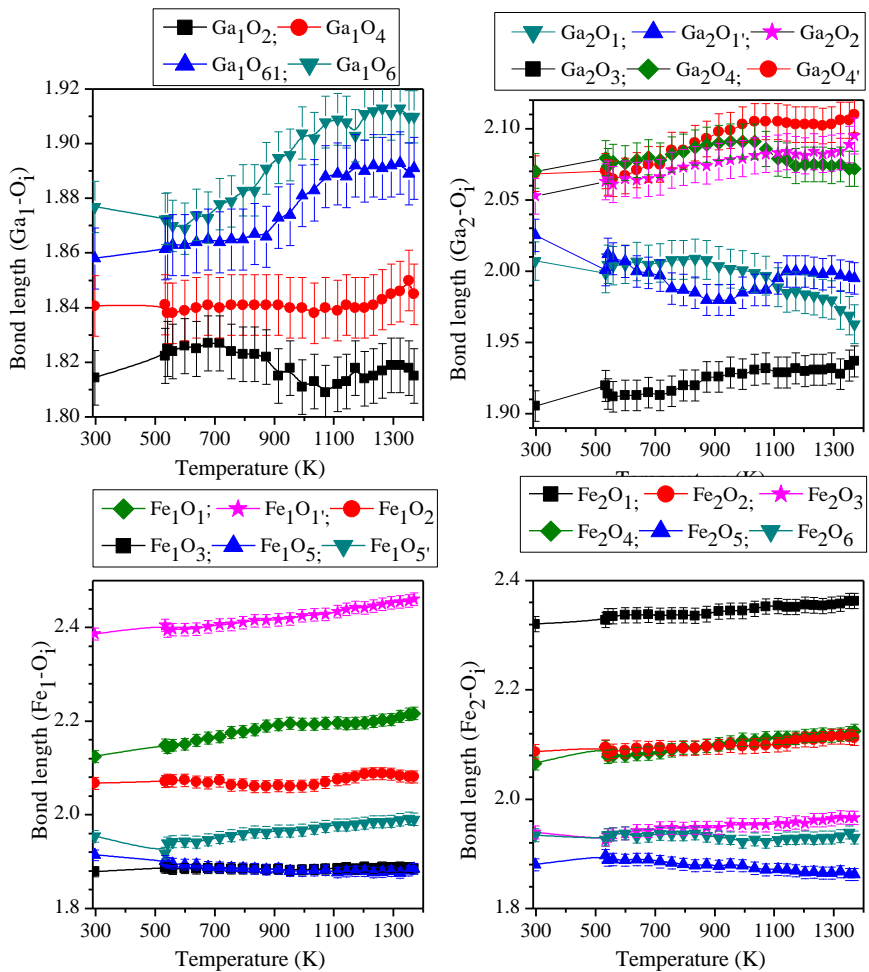

**Fig. 6** (color online) Evolution of the interatomic distance around the Ga$_1$, Ga$_2$, Fe$_1$ and Fe$_2$ cations as a function of temperature obtained after Rietveld refinement of GaFeO$_3$.

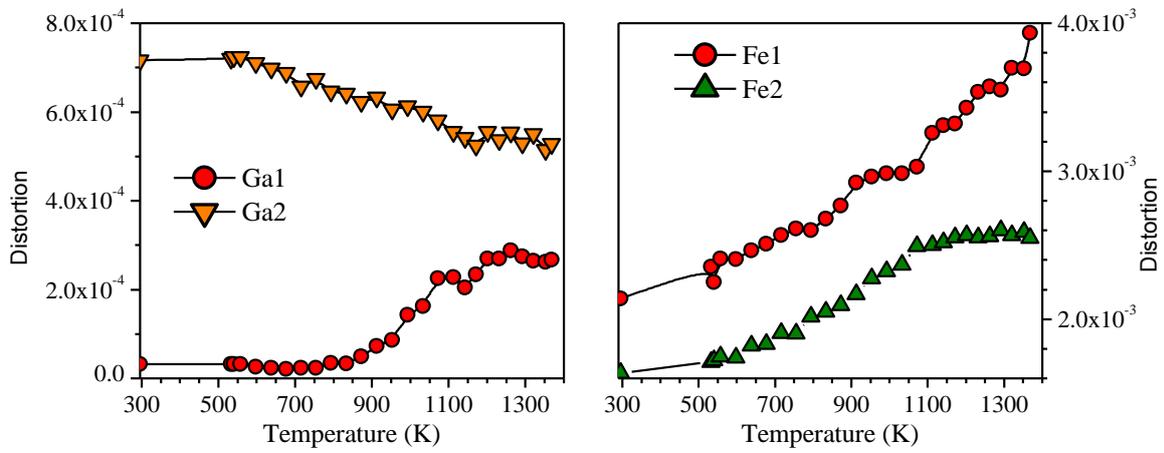

**Fig. 7** (color online) Evolution of the distortions of the oxygen polyhedra around the $Ga_1$, $Ga_2$, $Fe_1$ and $Fe_2$ cations.

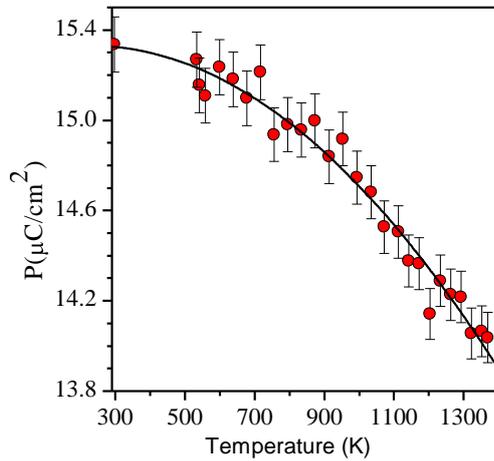

**Fig. 8** (color online) Evolution of the spontaneous polarization as a function of temperature calculated using simple ionic model described in the text.

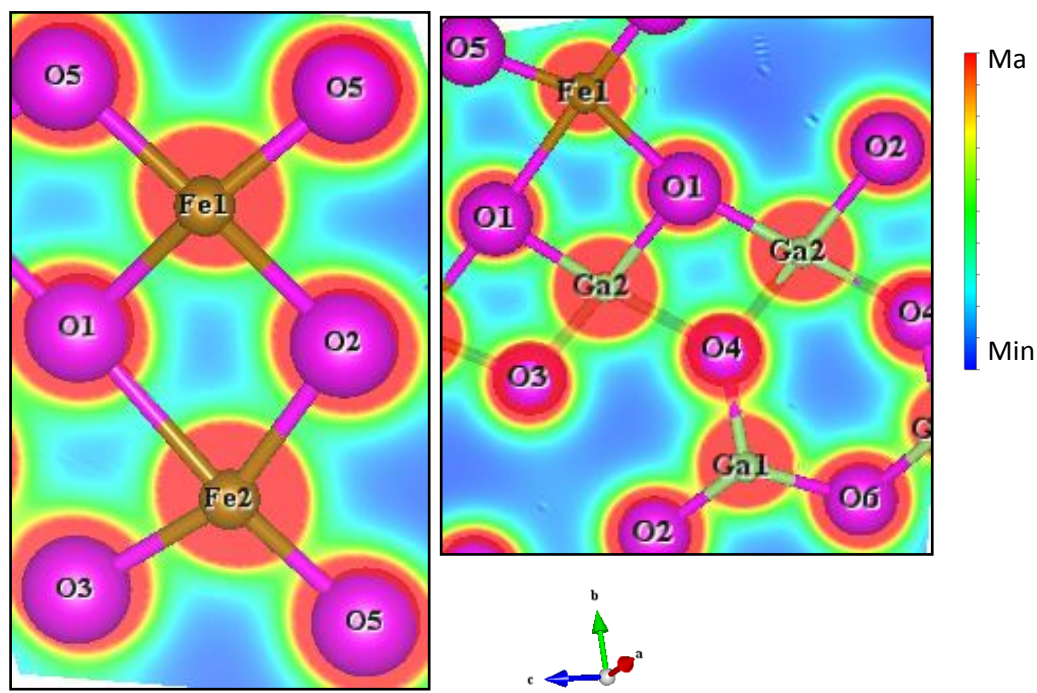

**Fig. 9** (color online) Computed charge density using VESTA software [32] and structural parameters obtained after Rietveld refinement at 296 K.